\documentclass[twocolumn,showpacs,preprintnumbers,amsmath,amssymb]{revtex4}
\usepackage{cases}
\usepackage{amssymb}
\usepackage{amsmath}
\usepackage{pifont}
\usepackage{graphicx}
\usepackage{dcolumn}
\usepackage{bm}
\usepackage{longtable}
\usepackage{hyperref}

\begin{document}
\title{\bf{Reply to ``Comment on `Semiquantum-key distribution using less than four quantum states' ''}}
\author{Xiangfu Zou$^{1,2}$}\email{xf.zou@hotmail.com (Xiangfu Zou);}
\author{Daowen Qiu$^{1}$}\email{issqdw@mail.sysu.edu.cn (Daowen Qiu).}
\affiliation{$^{1}$ Department of Computer Science, Sun Yat-sen University, Guangzhou 510006, China\\
$^{2}$School of Mathematics and Computational Science, Wuyi University,
Jiangmen 529020,  China}
\date{\today}

\begin{abstract}
Recently Boyer and Mor [\href{http://arxiv.org/abs/1010.2221}{arXiv:1010.2221 (2010)}] pointed out the first conclusion of Lemma 1 in our original paper [\href{http://link.aps.org/doi/10.1103/PhysRevA.79.052312}{Phys. Rev. A \textbf{79}, 052312 (2009)}] is not correct, and therefore, the proof of Theorem 5 based on Lemma 1 is wrong. Furthermore, they gave a direct proof for Theorem 5 and affirmed the conclusions in our original paper.  In this reply, we admit the first conclusion of Lemma 1 is not correct, but we need to point out the second conclusion of Lemma 1 is correct. Accordingly, all the proofs for Lemma 2, Lemma 3, and Theorems 3--6 are only based  on the the second conclusion of Lemma 1 and therefore are correct.
\end{abstract}
 \maketitle

The idea of semiquantum key distribution (SQKD) in which one of the parties (Bob) uses only classical operations was recently introduced \cite{BKM07}. Also, an SQKD protocol (BKM2007) using all four BB84 \cite{BB84} states was suggested \cite{BKM07}. Based on this, we presented some SQKD protocols which Alice sends less than four quantum states and proves them all being completely robust \cite{ZOU09}. In particular, we proposed two SQKD protocols in which Alice sends only one quantum state $|+\rangle$. Very recently, Boyer and Mor \cite{BM10} pointed out the first conclusion of Lemma 1 in our original paper \cite{ZOU09} is not correct, and therefore, the proof of Theorem 5 based on Lemma 1 is wrong. Furthermore, they gave a direct proof for Theorem 5 and affirmed the conclusions in Ref.\;\cite{ZOU09}.

In this reply, we first thank professors Boyer and Mor \cite{BM10} for their attention to our work and admit the first conclusion of Lemma 1 in Ref.\;\cite{ZOU09} is not correct. Particularly, we want to thank them for they not only pointed out the error in our paper but also gave a proof for Theorem 5 and confirmed the result of Theorem 5  in our original paper.

In this reply, we would also like to point out the second conclusion of Lemma 1 is correct. Accordingly, all the proofs for Lemma 2, Lemma 3, and Theorems 3--6 are only based  on the the second conclusion of Lemma 1 and therefore are correct.
To delete the first conclusion of Lemma 1 in Ref.\;\cite{ZOU09}, we only need to define the final combining state $\rho_i'^{AB}$ of Alice's $i$th particle and Bob's $i$th particle and modify Lemma 1 as follows.

{\bf Lemma 1.} Let $\rho'^{AB}$ denote Alice and Bob's final combining state and let $\rho_i'^{AB}$ be the final combining state of
Alice's $i$th particle and Bob's $i$th particle. If the attack $(U_E,U_F)$ induces no error on CTRL and TEST bits, then $\rho'^{AB}$ satisfies the following conditions:

(1) If $b_i=0$, then $\rho_i'^{AB}=(|\phi_i\rangle\langle\phi_i|)_A\otimes(|0\rangle\langle0|)_B$, i.e., Alice's $i$th final state is the sent state $|\phi_i\rangle$;

(2) If $b_i=1$, then $\rho_i'^{AB}= (x|00\rangle+y|11\rangle)(\overline{x}\langle
00|+\overline{y}\langle 11|)$ when the sent state $|\phi_i\rangle=x|0\rangle+y|1\rangle$,
i.e., the final combining state of Alice's $i$th particle and Bob's $i$th particle is the pure state $x|00\rangle+y|11\rangle$.

{\it Proof.} (1) The case of $ b_i  = 0 $.

The $i$th bit is a CTRL bit.   Alice's final quantum state $\rho_i'^A\ne |\phi _i
\rangle\langle\phi _i| $ can be detected by Alice as an error with
some non-zero probability. Also, Bob's $i$th final state is $|0\rangle$ since it is not acted any operation.
Thereby $\rho_i'^{AB}=(|\phi_i\rangle\langle\phi_i|)_A\otimes(|0\rangle\langle0|)_B$.

(2) The case of $ b_i  = 1 $.

The probability of the $i$th bit being a TEST bit is about
$\frac{1}{2}$. Also, if $|\phi_i\rangle=x|0\rangle+y|1\rangle$,
$\rho_i'^{AB}\ne (x|00\rangle+y|11\rangle)(\overline{x}\langle
00|+\overline{y}\langle 11|)$
 can be detected by Alice and Bob
 as an error with some non-zero probability when the $i$th bit is a TEST bit. Therefore
 $\rho_i'^{AB}= (x|00\rangle+y|11\rangle)(\overline{x}\langle
00|+\overline{y}\langle 11|)$.
\hfill{ } \ding{110}

The proof of Lemma 2 in Ref.\;\cite{ZOU09} is only  based on the second conclusion of Lemma 1 in Ref.\;\cite{ZOU09}. That is, Lemma 2 in Ref.\;\cite{ZOU09} also holds  when Lemma 1 is reformed as the above form. Because the proofs of Lemma 3 and Theorems 3--6 are only based  on Lemma 2 in Ref.\;\cite{ZOU09}, these results still hold.

\end{document}